\documentclass[twocolumn,showpacs,preprintnumbers,amsmath,amssymb,prl]{revtex4}
\usepackage{amsmath}

\usepackage{graphicx}
\usepackage{dcolumn}
\usepackage{bm}

\begin{document}

\title{High momentum entanglement of cold atoms generated by a single photon scattering}

\author{Rui Guo}
\author{Xiaoji Zhou}
\author{Xuzong Chen}\thanks{E-mail: xuzongchen@pku.edu.cn.}

\affiliation{School of Electronics Engineering and Computer Science,
Peking University, Beijing
100871, China}%

\date{\today}

\begin{abstract}
With the mechanism of pairwise scattering of photons between two
atoms, we propose a novel scheme to highly entangle the motional
state between two ultracold neutral atoms by a single photon
scattering and detection. Under certain conditions, it is shown that
an arbitrary amount of entanglement can be obtained with this
scheme.
\end{abstract}

.

\maketitle

\begin{figure}
\centering
\includegraphics[height=4.5 cm]{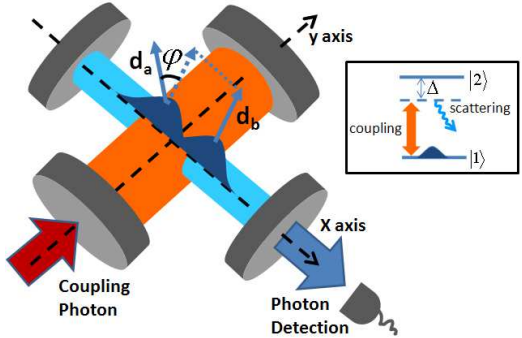}
\caption{(Color online) Schematic plot of the physical model. The
coupling photon is injected in a single--mode cavity along the
y-axis, whereas the scattering photon is collected and detected
along x-axis. The recoil of atoms is restricted in the x--direction.
The inset shows the atomic configuration. }
\end{figure}

\emph{Introduction.}--- Quantum entanglement, as one of the central
topics in quantum mechanics, has been extensively studied on the
Hilbert spaces with continuous variables (CV) in recent years
\cite{Accumulation of Entanglement,Anisotropically in Parametric
Down-Conversion,atom-atom entag. by detection,EPR,Parametric Down
Conversion,rmp,scattering,Schmidt num}. Compared to the finite
dimensional entanglement on photonic polarizations or atomic
internal states, the CV system provides entanglement in a rich
diversity of forms \cite{EPR,Anisotropically in Parametric
Down-Conversion,scattering,Parametric Down Conversion,atom-atom
entag. by detection}, which provide unique roles in quantum
information processing \cite{rmp,Accumulation of Entanglement} and
in fundamental test for quantum physics \cite{EPR}. It is only in CV
systems that the unbounded high degree of entanglement becomes
possible, which has been studied recently for optical squeezed state
\cite{rmp}, optical parametric down conversion \cite{Anisotropically
in Parametric Down-Conversion,Parametric Down Conversion},
atom--photon scattering \cite{scattering}, and atom--atom motional
state \cite{atom-atom entag. by detection}, etc..

For motional entanglement between neutral atoms without
interactions, it is verified that \cite{atom-atom entag. by
detection} one can not produce entanglement beyond $1$ ebit with
single-photon emission and detection, therefore, it unavoidably
requires nonlinear bi-photon or multi-modes detections for high
entanglement \cite{atom-atom entag. by detection}. By introducing
effective interactions between atoms in this letter, however, we
find that it is possible to produce arbitrarily high entanglement by
only a single photon scattering and detection, which is more
feasible and efficient for realistic implementations
\cite{Accumulation of Entanglement}. The basic physics for this
scheme is that, for a couple of cold atoms with the same electric
dipoles, photon scattered from one atom can be efficiently
re-absorbed and re-scattered by the other, this mechanism of
``pairwise scattering'' \cite{backward super radi,two light super
radi} of photons entangles the atomic wavepackets under the law of
momentum conservation. Evaluated by the Schmidt number $K$
\cite{Parametric Down Conversion,scattering}, it is shown that an
arbitrary amount of entanglement can be produced for ultracold atoms
with proper physical control parameters.

\emph{Theoretical analysis.}--- To describe the physical process, as
shown in Fig. 1, it is assumed that a pair of identical cold atoms
``a'' and ``b'' are coupled to a single photon in a cavity along
y-axis, with nearly parallel atomic electric dipoles $\vec{d}_{{\rm
a},{\rm b}}$, i.e., $|\vec{d}_{a}|\!=\!|\vec{d}_{b}|\!=\!d$ and
their relative angle $\varphi\approx 0$. We assume the coupling
photon is scattered and therefore recoils the atoms to the
x--direction where the photon detector is placed. Considering the
momentum exchange along x--direction, the Hamiltonian can be written
as $\hat{H}_{{\rm total}}=\hat{H}_{0}+\hat{H}_{{\rm I}}$, where
\begin{eqnarray}
\hat{H}_{0}&=&\frac{ (\hat{P}^{a}_{x})^{2}}{2m}+\frac{
(\hat{P}^{b}_{x})^{2}}{2m}+\sum_{k,s}\hbar
\omega_{k}a^{\dag}_{k,s}a_{k,s} \nonumber \\
&+&\hbar \omega_{21}(\sigma_{22}^{a}+\sigma_{22}^{b})
+\hbar \omega_{c}a^{\dag}a, \nonumber \\
\hat{H}_{{\rm I}}&=& \hbar
\sum_{k,s}\big{[}g_{a}(k,s)e^{-ikx_{a}}\sigma_{12}^{a}a^{\dag}_{k,s}\nonumber \\
&+&g_{b}(k,s)e^{-ikx_{b}}\sigma_{12}^{b}a^{\dag}_{k,s}+{\rm
H.C.}\big{]}
\nonumber \\
&+& \hbar \left( g_{c} \sigma_{12}^{a} a^{\dag}+g_{c}
\sigma_{12}^{b}a^{\dag}+{\rm H.C.} \right). \label{hamiltonian}
\end{eqnarray}
$\hat{P}^{a(b)}_{x}$ is the x-dimensional momentum operator for the
two-level atom ``a'' (``b'') with mass $m$ and transition frequency
$\omega_{21}$, and $\sigma_{ij}\equiv|i\rangle\langle j|$.
$a^{\dag}$ ($a$) is the creation (annihilation) operator for the
coupling single mode with frequency $\omega_{c}$ and coupling
strength $g_{c}$; whereas $a^{\dag}_{k,s}$ ($a_{k,s}$) is for the
scattering modes along x-axis with frequency $\omega_{k}=c|k|$ and
polarization $s$ and the coupling strength is
$g_{a(b)}(k,s)=\sqrt{\frac{c |k|}{2\varepsilon_{0}\hbar V}}\cdot
\vec{d}_{a(b)}\cdot \hat{\epsilon}_{k,s}$.

\begin{figure}
\centering
\includegraphics[height=3.8cm]{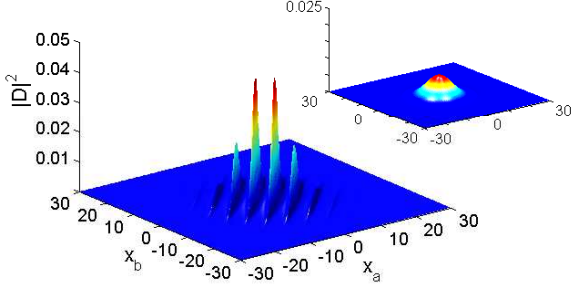}
\caption{(Color online) Spatial probability distribution of the
steady bipartite wavefunction $D(\mathbf{x},k\!=\!k_{c})$, with
$k_{c}\!=\!1$, $\sigma\!=\!0.2$, $\delta\!=\!0.1$ and $E_{m}\!=\!0$
\cite{explain in Er}. The small figure shows the initial Gaussian
distribution.}
\end{figure}

\begin{figure}
\centering
\includegraphics[height=4.5cm]{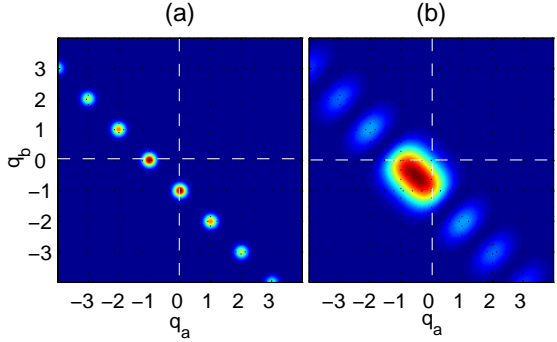}
\caption{(Color online) Density plot of momentum distributions for
the steady bipartite wavefunctions $D(\mathbf{q},k\!=\!k_{c})$ with
$k_{c}\!=\!1$, $\delta\!=\!0.1$, and $E_{m}\!=\!0$ \cite{explain in
Er}. (a) Different orders of scattered wavepackets are well
separated with $\sigma\!=\!0.3$. (b) Wavepackets begin to overlap
and interfere with $\sigma\!=\!1$.}
\end{figure}

We utilize the Dirac ket to denote the physical state, e.g.,
$|1_{a},q_{a};1_{b},q_{b};k\rangle $ represents that the atom ``a''
(``b'') has the internal state $|1\rangle_{a(b)}$ and momentum
$\hbar q_{a(b)}$, and the photon is scattered with momentum $\hbar
k$. In the slowly--varying frame, therefore, the system state can be
expanded as:
\begin{eqnarray}
&&| \Psi \rangle = \int\!\! {\rm d}\mathbf{q}\
e^{-i(E/\hbar+\omega_{c})t}\bigg[A(\mathbf{q})|1_{a},q_{a};2_{b},q_{b};0\rangle  \nonumber\\
&&
+B(\mathbf{q})|2_{a},q_{a};1_{b},q_{b};0\rangle+C(\mathbf{q})|1_{a},q_{a};1_{b},q_{b};0\rangle
\bigg ]\nonumber\\
&&+\int\!\! {\rm d}\mathbf{q}{\rm d}k\
e^{-i(E/\hbar+\omega_{k})t}D(\mathbf{q},k)|1_{a},q_{a};1_{b},q_{b};k\rangle,
\label{state expansion}
\end{eqnarray}
where the bold symbol $\mathbf{q}$ simplifies the pairwise variables
$(q_{a},q_{b})$, and the summation over the polarization $s$ is
implied in the integration over $k$. The kinetic energy is
$E\!\equiv\!\hbar^{2}(q_{a}^{2}+q_{b}^{2})/2m$.

In Eq. (\ref{state expansion}), $D(\mathbf{q},k)$ is the momentum
wavefunction for the atoms and the scattered photon, which is our
main concern for its induced atom--atom entanglement in the steady
state. $A(\mathbf{q})$ [$B(\mathbf{q})$] is the instantaneous
wavefunction for the atom ``b'' (``a'') is excited, and we introduce
$M(\mathbf{q})\!\equiv\! A(\mathbf{q})\!+\!B(\mathbf{q})$ for the
symmetry. From the Schr\"{o}dinger equation, the dynamical equations
read:
\begin{eqnarray}
\frac{{\rm d}M(\mathbf{q})}{{\rm d}t}&=&-i\Delta
M(\mathbf{q})-2ig_{c}^{*}C(\mathbf{q})-\gamma
M(\mathbf{q}) \nonumber \\
&-&e^{iE_{m}t/\hbar}\!\! \int \!\! {\rm d}k \!\!
\int_{0}^{t}\!\!{\rm d}s\ \big[ g_{a}(k)g_{b}(k)\times \nonumber \\
&&e^{i(\omega_{c}-c|k|)(t-s)}M(q_{a}\!+\!k,q_{b}\!-\!k,s)\big],
\label{dynamical equation 1} \\
\frac{{\rm d}C(\mathbf{q})}{{\rm d}t}&=&-ig_{c}M(\mathbf{q}),
\label{dynamical equation 2}\\
\frac{{\rm d}D(\mathbf{q},k)}{{\rm
d}t}&=&-i\exp{[-i(\omega_{c}-c|k|)t]}\times \label{dynamical equation 3} \\
&&\left[g_{b}(k)A(q_{a},q_{b}+k)+g_{a}(k)B(q_{a}+k,q_{b})\right],\nonumber
\end{eqnarray}
where the detuning $\Delta\!=\!\omega_{21}\!-\!\omega_{c}$ and
$\gamma=2\pi\int\!\!{\rm d}k\
|g_{a(b)}(k)|^{2}\delta(\omega_{k}\!-\!\omega_{c})$ is the atomic
natural linewidth, and $\Gamma\equiv 2|g_{c}|^{2}\gamma/\Delta^{2}$
denotes the scattering rate. $E_{m}\!=\!\hbar^{2}k_{c}^{2}/m$
represents the kinetic energy mismatch generated when the coupling
photon is scattered from one atom to another and be re-scattered
back to the coupling mode \cite{backward super radi,two light super
radi}. When this energy deficit is significant, the pairwise
scattering happens only in a time scale $\tau < \hbar/E_{m}$ and
further cascaded processes will be eliminated \cite{backward super
radi}. In this work, we assume $E_{m}$ is small enough to allow
needed cascaded pairwise scattering, and treat it as a constant
\cite{explain in Er}.

The fourth term at the r.h.s. of Eq. (\ref{dynamical equation 1})
represents the pairwise scattering between two atoms. If the atomic
dipoles are perpendicular ($\varphi\!=\!\pi/2$), this term vanishes
and the pairwise scattering will be forbidden in the Eqs.
(\ref{dynamical equation 1}--\ref{dynamical equation 3}), which
therefore reduce to the model of a single atom scattering
\cite{scattering}. When the dipoles are parallel
($\varphi\!\approx\!0$), however, photon scattered by one atom can
be efficiently re-scattered by the other, therefore produces
momentum correlation between them.

Eqs. (\ref{dynamical equation 1}--\ref{dynamical equation 3}) can be
analytically decoupled in the atomic position coordinates. We use
the argument $\mathbf{x}\!\equiv\! (x_{a},x_{b})$ for wavefunctions
to indicate their fourier counterparts, e.g., $M(\mathbf{x})\equiv
\int\!\!{\rm d}\mathbf{q}\
e^{-i\mathbf{x}\cdot\mathbf{q}}M(\mathbf{q})$, where
$\mathbf{x}\cdot\mathbf{q}\!\equiv\! x_{a}q_{a}\!+\! x_{b}q_{b}$.
With weakly coupling conditions: $g_{c},\gamma\ll\Delta$, and
$E_{m}/\hbar\ll\Gamma$, the adiabatic solution of Eq.
(\ref{dynamical equation 3}) yields:
\begin{eqnarray}
D(\mathbf{q},k,t)&=&N\!\! \int\!\!{\rm d}\mathbf{x}\
e^{i\mathbf{x}\cdot\mathbf{q}}
G(\mathbf{x})(e^{ix_{a}k}\!+\!e^{ix_{b}k})\times \nonumber\\
&&[1- e^{-t\cdot\Pi
(\mathbf{x},k,t)}]/\Pi (\mathbf{x},k,t), \label{timevarying solution}\\
\Pi(\mathbf{x},k,t)&=&i(\omega_{c}-2|g_{c}|^{2}/\Delta-c|k|)\nonumber \\
&+&\!\! \Gamma[1\!+{\rm
sinc}(E_{m}t/\hbar)\cos(\varphi)\cos{(k_{c}x_{a}\!-\!k_{c}x_{b})}],\nonumber
\end{eqnarray}
where $G(\mathbf{x})$ is the fourier counterpart of the initial
atomic momentum wavefunction
$G(\mathbf{q})\!=\!\sqrt{2}\exp{[-(q_{a}^{2}+q_{b}^{2})/\sigma^{2}]}/\sqrt{\pi}\sigma$
which is generally specified as an unentangled Gaussian with
momentum width $\sigma$. For atoms with nearly parallel dipoles
($\varphi\!\approx\! 0 $), we introduce $\cos (\varphi)\equiv \!1
\!-\! \delta^{2}$ and use $\delta$ $(\ll 1)$ as a dipole parallelity
parameter in the following. $N$ is the normalized factor and ${\rm
sinc}(x)\equiv\sin{(x)}/x$.

\emph{Generation of atom-atom entanglement with single-photon
detection}--- Eq. (\ref{timevarying solution}) is the time evolution
of collective atoms--photon wavefunction, which will be projected to
an atom--atom entangled state \cite{atom-atom entag. by detection}
after the detection of the scattered photon. In the following, we
neglect the light shift ($2|g_{c}|^{2}\!/\Delta$) and assume the
photon clicks on a narrow band detector with wavevector $k=+k_{c}$.

When $E_{m}\!\neq\! 0$, from Eq. (\ref{timevarying solution}), the
steady atomic momentum wavefunction reduces to a Bell-like state:
\begin{eqnarray}
D(\mathbf{q},k\!=\!k_{c},t\!\rightarrow\! \infty)\propto
G(q_{a}\!+\!k_{c},q_{b})+ G(q_{a},q_{b}\!+\!k_{c}),
\label{BellState}
\end{eqnarray}
since for $t\!\gg\! \hbar/E_{m}$ the energy--conserved
single--atom--scattering process prevails \cite{backward super
radi}, which can produce at most $1$ ebit by a single--photon
detection \cite{atom-atom entag. by detection}.

For $E_{m}\!=\!0$ \cite{explain in Er}, however, the atom--atom
pairwise scattering introduces effective interactions between them,
which produces correlated wavepacket as:
\begin{eqnarray}
&&\!\!\!\!\!\! D(\mathbf{q},k\!=\!k_{c},t\!\rightarrow\! \infty)\!\propto\! D'(q_{a}\!+\!k_{c},q_{b})\!+\! D'(q_{a},q_{b}\!+\!k_{c}), \label{momentum steadystate} \\
&&D'(\mathbf{q})=\exp{[-\frac{(q_{a}+q_{b})^{2}}{2\sigma^{2}}]}\exp{(-\frac{|q_{a}-q_{b}|\delta}{\sqrt{2}k_{c}})}\times\label{D'} \\
&&\sum_{n=0}^{\infty}\exp{[-\frac{\sigma^{2}\pi^{2}}{8k_{c}^{2}}(2n+1)^{2}]}\cos{[\frac{\pi}{2k_{c}}(2n\!+\!1)(q_{a}\!-\!
q_{b})]}.\nonumber
\end{eqnarray}
Eq. (\ref{D'}) characterizes the essences for the pairwise
scattering: the first Gaussian factor on the r.h.s. represents the
bipartite total momentum which is provided by the initial momentum
width $\sigma$; the Fourier series on the second line, which gives a
periodic structure of the relative momentum, characterizes different
orders of pairwise scattering, with the total number of scattering
orders controlled by the dipole parallelity parameter $\delta$
through the second Gaussian factor on the first line.

The spatial and momentum distributions of the atomic wavefunction of
Eq. (\ref{momentum steadystate}) are plotted in Fig. 2 and Fig. 3,
respectively. In Fig. 2, one sees that, the scattering of photon
with wavevector $k\!=\!k_{c}$ correlates the atomic wavepackets with
resonant spatial period $\lambda_{c}$, and produces grating-like
wavefunction with highly localized atomic relative positions.
Meanwhile, as in Fig. 3, the atomic momentum is correlated along
$q_{a}\!+\!q_{b}\!+\!k_{c}\!=\!0$ due to the momentum conservation.
When the initial atomic momentum width $\sigma$ is well below the
recoil momentum ($\sigma\! \ll\! k_{c}$) as in Fig. 3 (a), different
orders of scattered wavepackets are clearly separated, which
corresponds to a well entangled bipartite state; when $\sigma$ tends
towards $k_{c}$, however, as in Fig. 3 (b), recoiled wavepackets
begin to overlap and exhibit destructive quantum interferences for
the high--order scattering, which will eventually destroy the
momentum entanglement that can be seen more clearly in the
following.

The entanglement encoded in the wavefunction Eq. (\ref{momentum
steadystate}) can be quantitatively evaluated by the Schmidt
decomposition \cite{Schmidt num} of $D(\mathbf{q},k\!\!=\!\!k_{c})$:
\begin{eqnarray}
D(\mathbf{q},k\!\!=\!\!k_{c})=\!\!\!\sum_{n=-\infty}^{\infty}\!\!\!\sqrt{\lambda_{n}}S_{n}(\mathbf{q}),
\label{Schmidt Dec Formula}
\end{eqnarray}
where $S_{n}(\mathbf{q})$ is the momentum Schmidt basis which is
orthonormal and separable, with $\sum_{n} \lambda_{n}=1$. The degree
of entanglement is then defined as the Schmidt number
\cite{Parametric Down Conversion,scattering}: $K\equiv
1/\sum_{n}\lambda_{n}^{2}$.

For the single--atom--scattering state in Eq. (\ref{BellState}),
compared with Eq. (\ref{Schmidt Dec Formula}), one has at most $K=2$
which is only $1$ ebit \cite{atom-atom entag. by detection}. For the
pairwise--scattering--state in Eq. (\ref{momentum steadystate}),
however, from the numerical results of $K$ in Fig. 4 in dependence
of $\delta$ and $\sigma$, one sees that it may highly exceed the $1$
ebit limit \cite{atom-atom entag. by detection} for cold atoms
($\sigma\!<\!k_{c}$) with parallel dipoles ($\delta\!\approx\! 0$).

For atoms cooled well below the recoil temperature, i.e.,
$\sigma\!\ll\!k_{c}$, the Fourier series in Eq. (\ref{D'}) makes the
relative momentum highly localized in each single period, therefore,
the summation over spatial frequency can be well approximated by a
summation over discrete localized momentum modes. Along with Eqs.
(\ref{momentum steadystate}--\ref{Schmidt Dec Formula}), it
eventually yields the Schmidt decomposition as a summation of
different orders of scattered wavepackets:
\begin{eqnarray}
&&\!\!\!\!\! S_{n}(\mathbf{q})\!\!=\!\! (-1)^{|n|+1}{\rm Sgn}(n)G(q_{a}\!-\! nk_{c},q_{b}\!+\! nk_{c}\!+\!k_{c}), \label{scattering modes}\\
&& \lambda_{n}\!\!=\!\!N' \left(
e^{-\sqrt{2}\delta|n|}\!-\!e^{-\sqrt{2}\delta|n\!+\!1|} \right)^{2},
\label{schmidt eigenvalues}
\end{eqnarray}
where ${\rm Sgn}(n)$ is the signum function with value $+1$ for
$n\geq 0$ and $-1$ for $n<0$, and $N'$ is a normalization factor.
With Eq. (\ref{schmidt eigenvalues}), the degree of entanglement is
obtained:
\begin{eqnarray}
K\equiv 1/\sum_{n}\lambda_{n}^{2}=\sqrt{2}/\delta, \ \ \ (\delta\ll
1). \label{K}
\end{eqnarray}

Eq. (\ref{K}) fits well with the numerical results when
$\sigma\!<\!k_{c}/2$ as shown in Fig. 4. It indicates that, by
utilizing ultracold atom pair, such as in BEC, where both conditions
$\delta\!\approx\!0$ and $\sigma\!\ll\! k_{c}$ are well fulfilled,
arbitrarily high atom--atom entanglement may be produced by a single
photon scattering and detection, once the mismatch energy $E_{m}$ is
well compensated \cite{explain in Er}.

For hotter atoms with temperature approaches or exceeds the recoil
temperature, or equivalently, $\sigma\!\gtrsim\!k_{c}$, the
scattered momentum wavepackets $S_{n}(\mathbf{q})$ will overlap each
other and exhibit quantum interferences. From Eq. (\ref{scattering
modes}), one sees that different orders of recoiled wavepackets take
the same shape as the initial wavepacket $G(\mathbf{q})$, together
with an extra overall phase: $(-1)^{|n|+1}{\rm Sgn}(n)$. Explicitly,
the first two orders of scattered wavepackets $S_{0}(\mathbf{q})$
and $S_{-1}(\mathbf{q})$ take the same overall phase ``$-1$'', and
therefore exhibit constructive interference; for higher--order
wavepackets, however, $S_{n}(\mathbf{q})$' take opposite phases
``$\pm 1$'' interchangeably, which will induce destructively
interferences as in Fig. 3 (b). Due to this quantum interference,
the entanglement will be significantly decreased as in Fig. 4,
therefore, one can hardly produce high entanglement for hot atoms
with $\sigma\!\gg\! k_{c}$, even when the conditions
$\delta\!\approx\! 0$ and $E_{m}\!=\!0$ \cite{explain in Er} are
fulfilled.

The time evolution of the produced atomic entanglement can be
obtained from the general solution in Eq. (\ref{timevarying
solution}). As in Fig. 5, the steady entangled state is established
with a time $\tau$ which can be estimated as $\tau\!\sim\!
1/(\delta^{2}\Gamma)$. For $\hbar/E_{m}\!\gg\! \tau$ \cite{explain
in Er}, the maximal entanglement can be obtained by a single photon
detection with the coupling time $T$ satisfying $\tau\!\ll\!
T\!\ll\! \hbar/E_{m}$. For the general case of $E_{m}\neq 0$, the
produced entanglement will first increase due to the pairwise
scattering, and then decrease to a Bell--like state
($K\!\approx\!2$) when $T\!\gg\! \hbar/E_{m}$, since the
energy--conserved single--atom scattering process dominates in this
time scale \cite{backward super radi}.

\begin{figure}
\centering
\includegraphics[height=5cm]{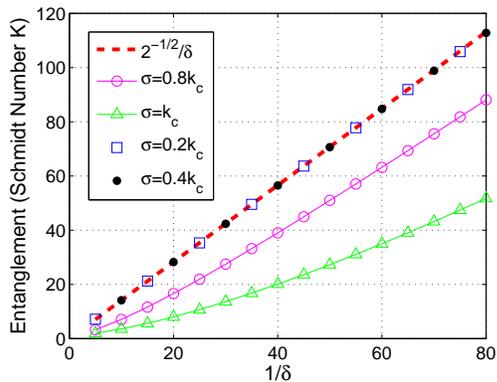}
\caption{(Color online) The Schmidt number $K$ for the entangled
steady state of $D(\mathbf{q},k\!=\!k_{c})$ is plotted against the
reciprocal dipole parallelity parameter $1/\delta$, with different
initial momentum widthes $\sigma$ as shown in the inset.}
\end{figure}

\begin{figure}
\centering
\includegraphics[height=4.5cm]{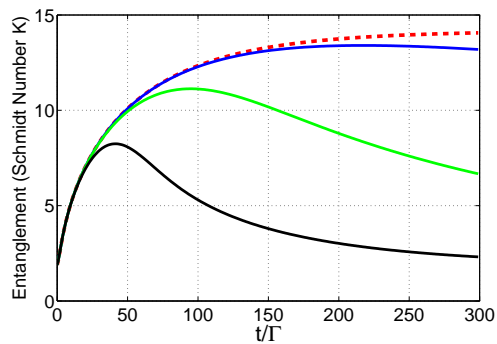}
\caption{(Color online) Time evolution of the produced entanglement.
The dashed red line indicates $E_{m}/\hbar\!=\!0$ \cite{explain in
Er}, whereas the solid lines from top to bottom are plotted with
$E_{m}/\hbar\!=\!0.1\!\times\!10^{-3}$ (blue),
$E_{m}/\hbar\!=\!0.5\!\times\!10^{-3}$ (green),
$E_{m}/\hbar\!=\!2\!\times\!10^{-3}$(black), respectively.
$\Gamma\!=\!1$, $\sigma\!=\!0.2k_{c}$, $\delta\!=\!0.1$.}
\end{figure}

\emph{Conclusion.---} From the first principle we have demonstrated
that, the mechanism of pairwise photon--scattering in a couple of
ultracold atoms may be used to produce superhigh atom--atom
entanglement. When the atom pair is cooled well below the recoil
temperature with parallel electric dipoles, it is possible to
achieve an arbitrary amount of entanglement by a single photon
detection once the scattering--induced energy mismatch $E_{m}$ is
compensated \cite{explain in Er}. This scheme can be used to produce
entangled atom resources for the test of EPR--nonlocality \cite{EPR}
and for quantum information processing \cite{rmp,Accumulation of
Entanglement}.

For experimental tests of this scheme, the spatially overlapping
ultracold atom pairs can be produced by coupling weak atom laser
beams from a BEC, and then be loaded into a cavity for detections of
the scattered photon \cite{Cavity exp}. Moreover, as in recent
report \cite{two light super radi} of BEC superradiant
\cite{backward super radi}, the generation of pairwise scattering by
atom pairs is preferred over single atom scattering when the energy
mismatch $E_{m}$ is compensated by the incident coupling light
\cite{two light super radi}, therefore, it is most probable to
analyze the quantum correlation in the pairwise scattering process
with this proposed model.

It is our pleasure to acknowledge the helpful comments on this
manuscript from Dr. Yu Ting. This work is supported by the state Key
Development Program for Basic Research of China (No.~2005CB724503
and 2006CB921401, 921402), NSFC(No. 60490280 and 10574005).


\begin{references}
\bibitem{explain in Er} $E_{m}$ can be made
smaller for heavier atoms; furthermore, it can be efficiently
compensated by extending the energy spectrum of the incident light,
see ref. \cite{two light super radi}. When $\hbar/E_{m}$ is much
larger than all other time scales in this physical process, we
simply set $E_{m}=0$.
\end{references}
\end{document}